\documentclass[conference]{IEEEtran}
\usepackage{cite}
\usepackage{amsmath,amssymb,amsfonts}
\usepackage{algorithmic}
\usepackage{graphicx}
\usepackage[caption=false]{subfig}
\usepackage{textcomp}
\usepackage{xcolor}
\usepackage{orcidlink}
\usepackage{transparent}
\usepackage{braket}
\usepackage{float}
\usepackage[linesnumbered, ruled,vlined]{algorithm2e}
\usepackage{amsmath}
\usepackage{tikz}
\usepackage{newtxtext}

\definecolor{lightcoral}{RGB}{240, 128, 128}
\definecolor{custompink}{RGB}{232,129,229}
\definecolor{customcyan}{HTML}{47ced0}
\definecolor{customgreen}{RGB}{80,205,60}

\def\rm#1{{\color{lightcoral} [RM: #1]}}

\def\BibTeX{{\rm B\kern-.05em{\sc i\kern-.025em b}\kern-.08em
    T\kern-.1667em\lower.7ex\hbox{E}\kern-.125emX}}
\begin{document}

\title{Quantum-Guided Cluster Algorithms for Combinatorial Optimization}

\makeatletter
\newcommand{\linebreakand}{%
  \end{@IEEEauthorhalign}
  \hfill\mbox{}\par
  \mbox{}\hfill\begin{@IEEEauthorhalign}
}
\makeatother

\author{
\IEEEauthorblockN{
\begin{minipage}{0.24\textwidth}
    \centering
    Peter J. Eder\textsuperscript{$\orcidlink{0009-0006-3244-875X}$} \\
    \textit{TUM, Garching, Germany} \\
    \textit{Siemens AG, Germany} \\
    peter-josef.eder@tum.de
\end{minipage}
\hfill
\begin{minipage}{0.24\textwidth}
    \centering
    Aron Kerschbaumer\textsuperscript{$\orcidlink{0009-0002-2370-8661}$}\\
    \textit{Institute of Science and Technology Austria (ISTA)} \\ aron.kerschbaumer@ist.ac.at
\end{minipage}
\hfill
\begin{minipage}{0.24\textwidth}
    \centering
    Jernej Rudi Fin\v{z}gar\textsuperscript{$\orcidlink{0009-0006-3244-875X}$} \\
    \textit{TUM Garching, Germany} \\
    \textit{BMW AG, Germany} \\
    jernej-rudi.finzgar@tum.de
\end{minipage}
\hfill
\begin{minipage}{0.24\textwidth}
    \centering
    Raimel A. Medina\textsuperscript{$\orcidlink{0000-0002-5383-2869}$}\\
    \textit{Institute of Science and Technology Austria (ISTA)} \\
    raimel.medina@ist.ac.at
\end{minipage}
}
\\[2ex] 
\IEEEauthorblockN{
\begin{minipage}{0.24\textwidth}
    \centering
    Martin J. A. Schuetz \\
    \textit{Amazon Advanced Solutions Lab, Seattle, Washington 98170, USA} \\
    maschuet@amazon.com
\end{minipage}
\hfill
\begin{minipage}{0.24\textwidth}
    \centering
    Helmut G. Katzgraber \\
    \textit{Amazon Advanced Solutions Lab, Seattle, Washington 98170, USA} \\
    helmut@katzgraber.org
\end{minipage}
\hfill
\begin{minipage}{0.24\textwidth}
    \centering
    Sarah Braun\textsuperscript{\orcidlink{0000-0002-7032-6116}} \\
    \textit{Siemens AG, Germany} \\
    sarah.braun@siemens.com
\end{minipage}
\hfill
\begin{minipage}{0.24\textwidth}
    \centering
    Christian B. Mendl\textsuperscript{\orcidlink{0000-0002-6386-0230}} \\
    \textit{TUM, Garching, Germany} \\
    christian.mendl@tum.de
\end{minipage}
}
}

\maketitle

\bstctlcite{BSTcontrol}

\begin{abstract}
Finding the ground state of Ising spin glasses is notoriously difficult due to disorder and frustration. Often, this challenge is framed as a combinatorial optimization problem, for which a common strategy employs simulated annealing, a Monte Carlo (MC)-based algorithm that updates spins one at a time. Yet, these localized updates can cause the system to become trapped in local minima. Cluster algorithms (CAs) were developed to address this limitation and have demonstrated considerable success in studying ferromagnetic systems; however, they tend to encounter percolation issues when applied to generic spin glasses.

In this work, we introduce a novel CA designed to tackle these challenges by leveraging precomputed two-point correlations, aiming solve combinatorial optimization problems in the form of \textsc{Max-Cut} more efficiently. In our approach, clusters are formed probabilistically based on these correlations. Various classical and quantum algorithms can be employed to generate correlations that embody information about the energy landscape of the problem. By utilizing this information, the algorithm aims to identify groups of spins whose simultaneous flipping induces large transitions in configuration space with high acceptance probability -- even at low energy levels -- thereby escaping local minima more effectively.

Notably, clusters generated using correlations from the Quantum Approximate Optimization Algorithm exhibit high acceptance rates at low temperatures. These acceptance rates often increase with circuit depth, accelerating the algorithm and enabling more efficient exploration of the solution space.
\end{abstract}


\section{Introduction}
\label{sec:Introduction}

Exploring Ising spin glasses poses significant computational difficulties due to their highly rugged energy landscapes caused by competing interactions (i.e., frustration) and disorder, thereby requiring efficient algorithms for accurately identifying low-energy states. Monte Carlo (MC) algorithms~\cite{Metropolis1953, Hastings1970, Geman1984, Tierney1994} offer a versatile framework for this task by stochastically sampling configurations to estimate statistical properties or locate low-energy states (see~\cite{Katzgraber2009IntroductionMC} for an introduction). At their core, MC methods, such as the Metropolis-Hastings algorithm, generate a Markov chain of states with transition probabilities designed to converge to the equilibrium distribution, making them suitable for studying thermodynamic properties. For optimization tasks, like finding ground states, MC can be adapted by guiding the sampling toward lower energies, though the rugged energy landscapes of spin glasses often limit their efficiency~\cite{Moreno2003, Wang2015}. Advanced variants, such as parallel tempering~\cite{Earl2005ParallelTempering} and population annealing~\cite{Weigel2017}, enhance the exploration of these landscapes by leveraging multiple replicas or populations, while simulated annealing (SA), an MC-inspired heuristic, focuses solely on optimization by gradually lowering a temperature parameter to approximate ground states. Despite these advances, the intrinsic complexity of spin glasses continues to hinder rapid convergence to ground states.

In contrast, ferromagnetic systems, where all spins interact in a way that favors uniform alignment, benefit from cluster MC algorithms, which accelerate convergence at low temperatures by updating groups of spins at once~\cite{Swendsen1987, Wolff1989}. Developing similar cluster methods for spin glasses could significantly enhance the performance of existing MC techniques, which could then be leveraged to efficiently solve combinatorial optimization problems. Notably, $\mathbb{Z}_2$-symmetric spin glasses can be directly transformed into \textsc{Max-Cut} instances, making them particularly important for combinatorial optimization, and thereby enabling a wide range of practical applications~\cite{Lucas2014}.

Swendsen and Wang~\cite{Swendsen1987} introduced a cluster algorithm (CA) that efficiently updates multiple spins in a single iteration by grouping them into clusters based on a probabilistic rule that ensures detailed balance, a condition that preserves the correct equilibrium distribution of the system. Wolff~\cite{Wolff1989} later proposed a variant that flips only a single cluster per iteration, making it particularly effective for ferromagnetic systems in any space dimension. Houdayer~\cite{Houdayer2001} introduced an algorithm that applies cluster moves between replicas at the same temperature while conserving total energy and maintaining detailed balance.

While the aforementioned CAs significantly accelerate simulations of ferromagnetic models, they struggle with generic spin-glass systems due to issues related to percolation~\cite{Radicchi2015}. In frustrated systems, where competing interactions prevent all spins from simultaneously satisfying their preferred alignments, percolation causes clusters to grow to sizes comparable to the system itself, limiting their effectiveness~\cite{Kessler1990}. For example, moves in Houdayer's algorithm require an underlying geometry with a percolation threshold  (i.e., the critical value of the occupation probability at which a large-scale connected cluster first emerges in the system) above 50\% to provide a significant speedup, which is the case for two-dimensional Ising spin-glass Hamiltonians but not for general spin glasses~\cite{Zhu2015}.

An alternative approach leverages self-organized criticality, using the Abelian sandpile model to generate power-law distributed avalanches and form clusters from affected spins~\cite{Hoffmann2018}. Unlike MC algorithms, this method does not rely on stochastic sampling but instead exploits emergent self-organized criticality. However, its effectiveness depends on the presence of self-organized critical behavior, which may not emerge in all problem structures. Additionally, the algorithm becomes less effective in highly connected graphs, where avalanche path-connectedness becomes irrelevant.

Recently, approaches leveraging quantum information, such as~\cite{Wurtz2024} \&~\cite{Fingar2024} or the quantum-enhanced Markov chain Monte Carlo algorithm in~\cite{Layden2023}, have been explored. For the latter, polynomial speedups over classical methods have been observed for small systems, but no Grover-type scaling advantage exists~\cite{Orfi2024}, and fine-tuned quenches are expected to be required for quantum speedups~\cite{Orfi2024quenches}.

In this paper, we address the aforementioned challenges by introducing a novel CA for finding ground states of generic spin glasses, leveraging precomputed classical or quantum two-point correlations to enhance optimization efficiency. By utilizing this information to form clusters, our approach aims to induce large transitions in configuration space -- even at very low energy levels -- enabling the algorithm to escape local minima more effectively compared to single-spin updates. This mechanism is illustrated in Fig.~\ref{fig:qgca_single_spin_vs_cluster}. Note that the CA can also be seen as a post-processing strategy, further refining the outputs of quantum optimization algorithms to achieve even lower-energy states or potentially optimal solutions.

\begin{figure}[htbp]
    \def\svgwidth{0.48\textwidth}
\begingroup%
  \makeatletter%
  \providecommand\color[2][]{%
    \errmessage{(Inkscape) Color is used for the text in Inkscape, but the package 'color.sty' is not loaded}%
    \renewcommand\color[2][]{}%
  }%
  \providecommand\transparent[1]{%
    \errmessage{(Inkscape) Transparency is used (non-zero) for the text in Inkscape, but the package 'transparent.sty' is not loaded}%
    \renewcommand\transparent[1]{}%
  }%
  \providecommand\rotatebox[2]{#2}%
  \newcommand*\fsize{\dimexpr\f@size pt\relax}%
  \newcommand*\lineheight[1]{\fontsize{\fsize}{#1\fsize}\selectfont}%
  \ifx\svgwidth\undefined%
    \setlength{\unitlength}{455.12399749bp}%
    \ifx\svgscale\undefined%
      \relax%
    \else%
      \setlength{\unitlength}{\unitlength * \real{\svgscale}}%
    \fi%
  \else%
    \setlength{\unitlength}{\svgwidth}%
  \fi%
  \global\let\svgwidth\undefined%
  \global\let\svgscale\undefined%
  \makeatother%
  \begin{picture}(1,0.64401873)%
    \lineheight{1}%
    \setlength\tabcolsep{0pt}%
    \put(0,0){\includegraphics[width=\unitlength,page=1]{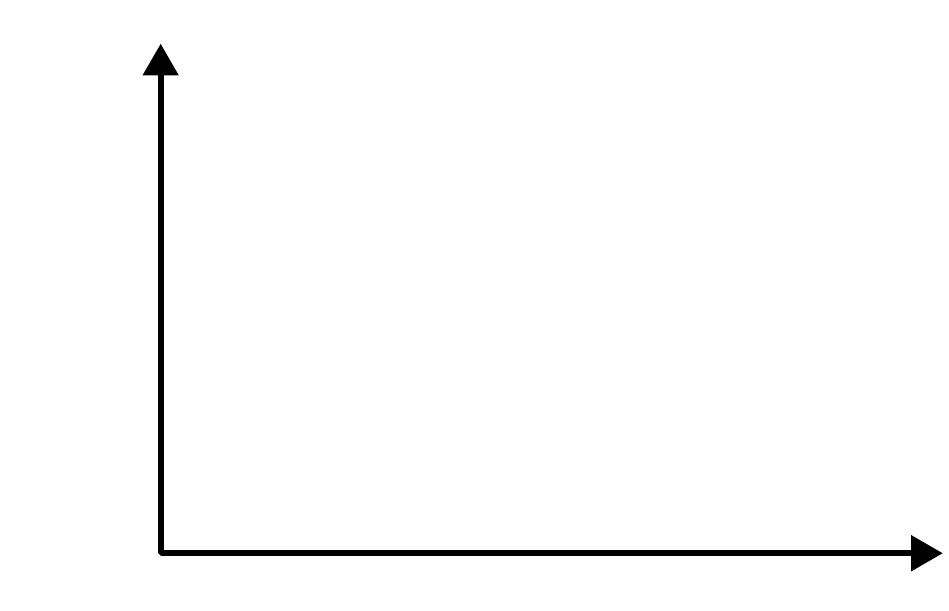}}%
    \put(0.13050061,0.61845908){\color[rgb]{0,0,0}\makebox(0,0)[t]{\lineheight{1.25}\smash{\begin{tabular}[t]{c}Energy\end{tabular}}}}%
    \put(0.99138972,0.00053214){\color[rgb]{0,0,0}\makebox(0,0)[t]{\lineheight{1.25}\smash{\begin{tabular}[t]{c}States\end{tabular}}}}%
    \put(0,0){\includegraphics[width=\unitlength,page=2]{qgca_single_spin_vs_cluster.pdf}}%
    \put(0.45005085,0.120929){\color[rgb]{0.17647059,0.80784314,0.09019608}\rotatebox{-4.9974896}{\makebox(0,0)[t]{\lineheight{1.25}\smash{\begin{tabular}[t]{c}\textbf{Cluster flip}\end{tabular}}}}}%
    \put(0,0){\includegraphics[width=\unitlength,page=3]{qgca_single_spin_vs_cluster.pdf}}%
    \put(0.0676874,0.36222318){\color[rgb]{0.60392157,0.25882353,0.23921569}\makebox(0,0)[t]{\lineheight{1.25}\smash{\begin{tabular}[t]{c}\textbf{Single-}\\\textbf{spin flip}\end{tabular}}}}%
    \put(0,0){\includegraphics[width=\unitlength,page=4]{qgca_single_spin_vs_cluster.pdf}}%
  \end{picture}%
\endgroup%

    \caption{A rugged energy landscape, characteristic of spin glasses, is depicted, along with illustrative transitions from single-spin-flip and cluster-flip updates.}
    \label{fig:qgca_single_spin_vs_cluster}
\end{figure}

Our contributions are as follows:
\begin{itemize}
    \item We introduce a novel MC CA that leverages precomputed information in the form of two-point correlations to enhance the performance of solving combinatorial optimization problems.
    \item Our results show that the algorithm with coupling constants outperforms SA on 3-regular graphs, with additional correlation information further improving performance, particularly on 20-degree graphs with higher levels of frustration.
    \item We analyze the mechanisms behind this improvement, providing insights into how frustration influences correlation-driven cluster formation and enhances optimization efficiency.
    \item By incorporating low-energy solutions from the Quantum Approximate Optimization Algorithm (QAOA), we demonstrate that quantum correlations accelerate the algorithm, enabling efficient cluster updates and improving exploration of the solution space.
\end{itemize}

\section{Methods}
\label{sec:Methods}

In this section, we first introduce the \textsc{Max-Cut} problem, then we explain the design of the proposed algorithm. Finally, we outline how the different correlation types that guide the CA are computed.

\subsection{The \textsc{Max-Cut} Problem \& Ising Spin Glasses}

The \textsc{Max-Cut} problem is a combinatorial optimization problem that is NP-hard~\cite{Karp1972}, such that finding optimal solutions for large, dense graphs with hundreds of nodes can already exceed the capabilities of state-of-the-art classical algorithms~\cite{Charfreitag2022}. Since \textsc{Max-Cut} formulations capture a wide range of NP-hard problems, this mapping naturally extends to real-world applications, including, for instance, manufacturing scheduling, and logistics planning~\cite{Lucas2014}.

A weighted \textsc{Max-Cut} instance is defined on a weighted, undirected graph $G = (V, E)$, where $V$ is the set of $n$ vertices and $E$ is the set of edges, each assigned a weight $A_{ij} \in \mathbb{R}$. The goal is to partition $V$ into two subsets such that the total weight of edges crossing the partition is maximized.

The corresponding cost function is given by:
\begin{equation}
C(\mathbf{x}) = \frac{1}{2} \sum_{\{i,j\} \in E} A_{ij} (1 - x_i x_j),
\label{eq:max_cut_objective_function}
\end{equation}
where $x_i \in \{-1,1\}$ represents the subset assignment of vertex~$i$.
This formulation can be mapped to the Hamiltonian of an Ising spin glass, given by 
\begin{equation}
H(\mathbf{x}) = - \sum_{\{i,j\} \in E} J_{ij} x_i x_j.
\label{eq:spin_glass}
\end{equation}
Here, the coupling constants $J_{ij}$ represent interaction strengths between spins. Finding the ground state of $H$ is mathematically equivalent to maximizing~\eqref{eq:max_cut_objective_function}. It is important to note here that our algorithm can easily be applied to generic spin glasses (i.e., also constrained optimization problems) by simply adding a magnetic field term to the Hamiltonian.

\subsection{Algorithmic Design}
\label{sec:Algorithmic_Design}

The proposed CA follows a structure similar to SA, with the key difference that, instead of flipping a single spin, an entire cluster of spins is formed and flipped at each step. A high-level outline is provided in Fig.~\ref{fig:algorithm_high_level} and the detailed pseudocode is given in Algorithm~\ref{alg:qcga_linear_schedule}, which we also denote as correlation-guided CA, or when employing quantum correlations, quantum-guided CA.

Before the algorithm begins, a correlation matrix $Z$ is computed, serving as the foundation of our approach by guiding the probabilistic formation of clusters and enabling efficient exploration of the solution space. Specifically, two-point correlations $Z_{i\neq j} \in [-1,1]$ indicate the likelihood of two spins $i$ \& $j$ belonging to the same partition, with values near $1$ suggesting alignment and values near $-1$ implying opposite partitions. Note that the computation of $Z$ is performed only once for the entire process. Section~\ref{sec:Deriving_the_Correlations} provides a detailed explanation of how the various types of correlations analyzed in this paper are computed. Importantly, the algorithm is not tailored to these specific correlation types, and the sources used can be replaced with any other classical or quantum alternative.

The algorithm starts by initializing a random spin configuration, $\mathbf{x}$. Then, in each iteration, a random seed node is selected, and a cluster is formed based on $Z$. The cluster building mechanism is explained in detail later in this section and illustrated in Fig.~\ref{fig:cluster_building}. The cluster $\mathcal{C}$ is then flipped (i.e., $\forall x_i \in \mathcal{C}: x_i \leftarrow -x_i$), the energy difference is computed, and the update is accepted or rejected according to the Metropolis criterion (see Algorithm~\ref{alg:qcga_linear_schedule} for details). The temperature is gradually reduced in each iteration to facilitate convergence, continuing until the final temperature $T_f = 1/\beta_f$ is reached.  

\begin{figure}[b]
    \def\svgwidth{0.48\textwidth}
    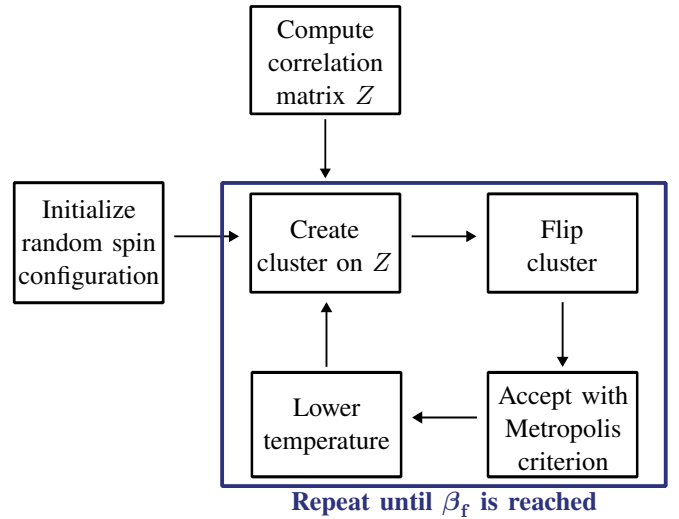
    \caption{A high-level scheme of the presented cluster algorithm, showing the individual steps performed in every iteration until the final inverse temperature $\beta_f$ is reached. Note that the correlations have to be calculated only once at the beginning.}
    \label{fig:algorithm_high_level}
\end{figure}

\begin{algorithm}[htbp]
\caption{Correlation-Guided Cluster Algorithm}
\label{alg:qcga_linear_schedule}
\SetAlgoLined
\KwIn{Hamiltonian $H$, correlation matrix $Z$, final inverse temperature $\beta_f$, number of iterations $m$, and probability scaling factor $\lambda_{\text{scale}}$.}
\KwOut{Best energy $E_{\text{best}}$ \& corresponding state vector $\mathbf{x}$.}

$\beta \leftarrow 0$\;
Initialize random spin configuration $\mathbf{x}$\;
$E_{\text{best}} \leftarrow H(\mathbf{x})$\;
\While{$\beta < \beta_f$}{
    $i \leftarrow$ Uniform$(1, n)$; \tcp*[f]{Choose seed node}\\
    $\mathcal{C} \gets$ CreateCluster$(i, \mathbf{x}, H, Z, \lambda_{\text{scale}})$\;
    $\mathbf{x}' \gets$ Flip$(\mathbf{x}, \mathcal{C})$; \tcp*[f]{Flip Cluster}\\
    $\Delta E \leftarrow H(\mathbf{x}') - H(\mathbf{x})$\;
    \If{$\Delta E \leq 0$}{
        $\mathbf{x} \leftarrow \mathbf{x}'$\;
        \If{$H(\mathbf{x}) < E_{\text{best}}$}{
            $E_{\text{best}} \leftarrow H(\mathbf{x})$\;
        }
    }
    \ElseIf{$\mathrm{Uniform}(0, 1) < e^{-\beta \Delta E}$}{
        $\mathbf{x} \leftarrow \mathbf{x}'$; \tcp*[f]{Metropolis criterion}
    }
    $\Delta \beta \leftarrow \beta_f/(m - 1) \text{len}(\mathcal{C})$\;
    $\beta \leftarrow \beta + \Delta \beta$; \tcp*[f]{Reduce temperature}\\
}
\end{algorithm}

\begin{figure*}[htbp]
    \centering
    \def\svgwidth{\textwidth}
    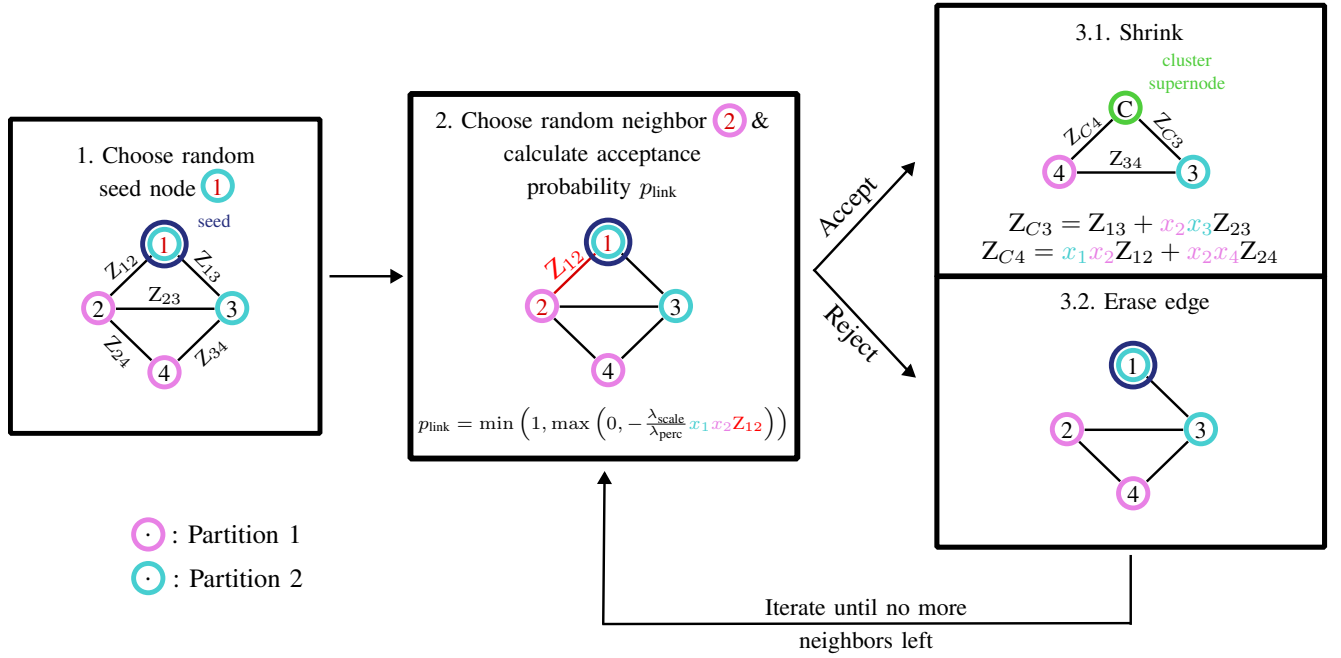
    \caption{Illustration of the cluster-building process, characterizing the pseudocode function \texttt{CreateCluster()}. Starting from a randomly selected seed node (here, $i=1$), neighboring vertices are iteratively considered for inclusion in the cluster with probability $p_{\text{link}}$, determined by the correlation matrix $Z$. Accepted nodes undergo a shrinking step before further expansion, while rejected nodes have their edge to the cluster removed.}
    \label{fig:cluster_building}
\end{figure*}

The schedule is designed to reach $\beta_f$ after $m$ iterations, where each iteration corresponds to flipping a single spin. Consequently, flipping a cluster of size $k$ is treated as equivalent to performing $k$ single-spin updates in SA. Note that while SA computes the exponential acceptance probability each time a spin flip increases the energy, our approach does so only once per cluster flip when the energy increases. This introduces a slight disadvantage for our method, but we adopt this approach for simplicity.

The cluster formation process in \texttt{CreateCluster()}, as detailed in Fig.~\ref{fig:cluster_building}, is guided by the correlation matrix $Z$. As mentioned in Section~\ref{sec:Introduction}, current CAs suffer from percolation issues when applied to spin glasses. By utilizing this precomputed information in form of $Z$, we aim to overcome this challenge.

To construct a cluster, the algorithm starts from a randomly chosen seed node and iterates over its neighbors in random order, adding each to the cluster with the link probability
\begin{equation}
p_{\text{link}} = \min\left(1,\max\left(0, -\frac{\lambda_{\text{scale}}}{\lambda_{\text{perc}}} x_i x_j Z_{ij}\right)\right).
\label{eq:p_link}
\end{equation}
Here, $\lambda_{\text{scale}}$ is a scaling factor that has to be tuned and
\begin{equation}
    \lambda_{\text{perc}} = \frac{\langle d \rangle}{2 \mathbb{E}[|Z_{i\neq j}| \mid Z_{i\neq j} \neq 0] \cdot (\langle d^2 \rangle - \langle d \rangle)}
    \label{eq:p_perc}
\end{equation}
is an estimate for the percolation threshold of the underlying graph, normalizing $p_{\text{link}}$ and thereby ensuring that clusters do not percolate the system. $\langle d \rangle$ and $\langle d^2 \rangle$ denote the first and second moments of the graph's degree distribution, respectively -- that is, the average node degree and the average of the squared degrees (which reflects the degree variability). $\mathbb{E}[|Z_{ij}| \mid Z_{ij} \neq 0]$ denotes the average absolute value of the non-zero entries of $Z$. This estimate is based on~\cite{Radicchi2015}, where it was shown to accurately predict the percolation threshold across various networks. We extend it by introducing the additional factor $1/(2 \mathbb{E}[|Z_{i\neq j}| \mid Z_{i\neq j} \neq 0])$  to account for the magnitude of the correlation matrix entries.

The neighboring node is either rejected -- causing the edge to be erased -- or accepted, in which case the newly added node and the seed node are shrunken into a new supernode representing the cluster, as detailed in Fig.~\ref{fig:cluster_building}. For more details on the shrinking process, see~\cite{Bravyi2020, Fischer2024}. The process continues until no further neighbors remain.

Finally, note that the algorithm can easily be extended to directly solve general spin glasses with magnetic field terms, by simply adjusting the Hamiltonian $H(\mathbf{x})$.

\subsection{Deriving the Correlations}
\label{sec:Deriving_the_Correlations}

In the proposed correlation-guided CA, the clusters are built based on precomputed information, such as coupling constants or low-energy correlations. In the following, we describe all correlation types that are considered in this paper and the methods used to obtain them.

\subsubsection{Coupling Constants \& Random Clusters}
\label{sec:Coupling_Constants} 

Coupling constants correspond directly to the problem interaction matrix $Z_{ij}^{\text{CC}} = J_{ij}$. As they provide no additional information beyond the problem structure, the CA is guided solely by the topology of the graph in this case. Random Clusters are generated using a pre-optimized, constant value for the probability $p_{\text{link}}$ of adding a spin to the cluster, as specified later in Section~\ref{sec:coupling_constants_and_random_clusters}.

\subsubsection{Semidefinite Programming}
\label{sec:Semidefinite_Programming}

To generate correlation matrices from Semidefinite Programming (SDP), we build on the Goemans and Williamson (GW) approximation algorithm~\cite{Goemans1995}. This type of correlation has previously been utilized in~\cite{Fischer2024} for shrinking algorithms. In the SDP relaxation, discrete spin variables are replaced by unit vectors on a unit sphere, transforming \textsc{Max-Cut} into a continuous optimization problem. This relaxed problem can be solved in polynomial time, producing a positive semidefinite matrix that encodes inner products of the solution vectors. The vectors ${\mathbf{v}_i}$ are then extracted via Cholesky decomposition, and the SDP correlations can be computed as $Z_{ij}^{\text{GW}} = \mathbf{v}_i^\text{T} \mathbf{v}_j$.

Importantly, these correlations reflect the probability that an edge $\{i,j\}$ is cut when applying the GW rounding procedure~\cite{Goemans1995}, where a random hyperplane partitions the solution vectors. On average, it achieves at least an approximation ratio of $C_{\text{app}}^{\text{GW}}=C(\mathbf{x})/C(\mathbf{x_{\text{opt}}})\simeq 0.878$, where $\mathbf{x}_{\text{opt}}$ denotes an optimal solution. SDP correlations are well-suited for our algorithm, as they can be computed in polynomial time.

\subsubsection{Thermal Correlations Sampled from Monte Carlo}
\label{sec:Monte_Carlo}

As will be discussed in Section~\ref{sec:Results}, we use MC correlations to validate our approach on larger graph sizes beyond the reach of QAOA simulations. This analysis also reveals how frustration influences behavior across graphs of varying degrees and how this frustration-driven behavior shifts when considering samples of different solution qualities.

To obtain thermal correlations via MC sampling (see~\cite{Katzgraber2009IntroductionMC} for an introduction), we use the Metropolis-Hastings algorithm~\cite{Metropolis1953} to generate spin configurations at various target temperatures $T_s = 1/\beta_s$. This process produces bitstrings of varying solution quality, with lower temperatures (higher $\beta_s$) biasing the system toward lower-energy states.

To ensure equilibration, we monitor energy and magnetization $M = \frac{1}{n} \sum_{i=1}^{n} x_i$. Given the $\mathbb{Z}_2$ symmetry of the model, $M$ averages to zero, with fluctuations within $1\%$ across all sampled temperatures. 

Finally, the spin-spin correlations are computed as:
\begin{equation}  
Z_{ij}^{\text{MC}}(\beta_s) = \left\langle x_i x_j \right\rangle_{\beta_s}.
\end{equation}

\subsubsection{Quantum Approximate Optimization Algorithm}
\label{sec:QAOA}

QAOA is a hybrid quantum-classical algorithm designed to approximate solutions to combinatorial optimization problems, including spin glasses~\cite{Farhi2014}. It mimics quantum adiabatic evolution, where the system evolves slowly enough to remain in its instantaneous ground state~\cite{Farhi2000, Born1928}.

In QAOA, the evolution begins from the uniform superposition state
\begin{equation}
\ket{+}^{\otimes n} = \frac{1}{\sqrt{2^n}} \sum_{\mathbf{x}} \ket{\mathbf{x}},    
\end{equation}
and proceeds through a sequence of alternating unitary transformations governed by the cost Hamiltonian $H_C$ and the mixer Hamiltonian $H_M = -\sum_{i=1}^{n} \sigma_i^x$ to prepare the state
\begin{equation}
\ket{\Psi(\boldsymbol{\beta}, \boldsymbol{\gamma})}=U_M(\beta_p)U_C(\gamma_p) \dots U_M(\beta_1)U_C(\gamma_1)\ket{+}^{\otimes n},
\end{equation}
where $U_C(\gamma_i) = e^{-i\gamma_i H_C}$ and $U_M(\beta_i) = e^{-i\beta_i H_M}$. In the case of $\mathbb{Z}_2$-symmetric Ising spin glasses, the cost Hamiltonian is given as:
\begin{equation}
    H_C=- \sum_{\{i,j\} \in E} J_{ij} \sigma^z_i \sigma^z_j.
\end{equation}
Here, $\sigma^x_i$ and $\sigma^z_i$ denote the Pauli-$X$ and Pauli-$Z$ matrices applied to the $i$-th qubit, respectively. The parameters $\boldsymbol{\beta}=(\beta_1, \ldots, \beta_p)$ and $\boldsymbol{\gamma}=(\gamma_1, \ldots, \gamma_p)$ with $p \in \mathbb{N}$ control the evolution and are optimized to minimize the expectation value $\bra{\Psi(\boldsymbol{\beta}, \boldsymbol{\gamma})} H_C \ket{\Psi(\boldsymbol{\beta}, \boldsymbol{\gamma})}$.  Finally, the two-point spin-spin correlations between vertices $i$ and $j$ can be written as
\begin{equation}
    Z_{ij}^{\text{QAOA}}=\bra{\Psi(\boldsymbol{\beta}_{\text{opt}}, \boldsymbol{\gamma}_{\text{opt}})} \sigma_i^z \sigma_j^z \ket{\Psi(\boldsymbol{\beta}_{\text{opt}}, \boldsymbol{\gamma}_{\text{opt}})},
\end{equation}
where $(\boldsymbol{\beta}_{\text{opt}}, \boldsymbol{\gamma}_{\text{opt}})$ are the optimized parameters.

Since QAOA is designed to provide low-energy approximate solutions to optimization problems, it is well-suited for the proposed CA, which serves as a post-processing strategy to refine solutions or even find the optimal one. A key feature of QAOA in deriving correlations is that the computationally expensive parameter optimization is performed only once. After training, the optimized parameters can be employed to efficiently sample high-quality solutions without further costly optimization. Finally, note that the quantum-guided CA is not exclusive to QAOA and can be executed with any alternative quantum optimization algorithm.

\section{Results}
\label{sec:Results}

In this section, we present the results of the CA, guided by the correlations introduced in Section~\ref{sec:Deriving_the_Correlations}. We begin by benchmarking the correlation-guided CA using coupling constants and random clusters in Section~\ref{sec:coupling_constants_and_random_clusters}, comparing its performance with that of SA. Next, in Section~\ref{sec:semidefinite_programming_and_monte_carlo_correlations}, we employ SDP correlations and thermal correlations sampled with the Metropolis-Hastings algorithm to demonstrate how the graph structure -- and, in particular, frustration -- affects the CA's performance when guided by correlations of varying quality. These analyzes set the stage for the main results presented in Section~\ref{sec:Correlations_Computed_from_QAOA}, where the CA guided by QAOA correlations is discussed.

In our experiments, all graph weights are chosen uniformly at random from $\{-1, 1\}$, and the final inverse temperature for both the correlation-guided CA and SA runs is set to $\beta_f=8$. The graph instances of size $n=28$ and the $3$-regular graphs of size $n=100$ were solved to optimality using Gurobi~\cite{gurobi}, while the $20$-regular graphs of size $n=100$ were solved using SA with $10{,}000n$ iterations and $50$ repetitions. Consequently, the approximation ratios $\bar{C}_{\text{app}}$ for the small instances and $3$-regular graphs are computed relative to guaranteed optimal solutions, whereas for the $20$-regular graphs with $n=100$, they are measured against the best solutions found.

\subsection{Coupling Constants \& Random Clusters}

\label{sec:coupling_constants_and_random_clusters}
In this section, we evaluate the performance of the correlation-guided CA using coupling constants and random clusters, comparing it with SA. We present the percentage of optimal solutions found, along with the standard deviations, across $20$ graph instances each of $3$-regular and $20$-regular graphs with $n=100$ nodes. The results are plotted in Fig.~\ref{fig:CC_RC_SA} as a function of iterations. For each graph and data point, the CA is run $100$ times to ensure statistical robustness.

For coupling constants, we choose $\lambda_{\text{scale}} = 1$, as that is where the correlation-guided CA achieves peak performance, which was identified via a preliminary optimization of $\lambda_{\text{scale}}$ over a range of values between $0.1$ and $10$ over $100n$ iterations, with $100$ repetitions per graph per value. This demonstrates that normalizing $p_{\text{link}}$ relative to the percolation threshold $\lambda_{\text{perc}}$ and applying our cost calculation method produces potentially meaningful clusters that do not span the entire system. A secondary, smaller peak in performance occurs at $\lambda_{\text{scale}} = 7.5$, which we attribute to the $\mathbb{Z}_2$ symmetry of Ising spin glasses. For higher values (e.g., $\lambda_{\text{scale}} = 10.0$), performance drops sharply due to percolation effects.

For random clusters, we set the link probability $p_{\mathrm{link}}=0.2$, determined through a preceding optimization process that tests $p_{\mathrm{link}} \in \{ 0.1, 0.2, \ldots, 0.9\}$ over $100n$ iterations and the same repetition protocol as above.

For the $3$-regular graphs, the correlation-guided CAs with coupling constants and random clusters both outperform SA, with coupling constants slightly surpassing random clusters. In contrast, on $20$-regular graphs, random clusters do not produce any optimal solution after $10{,}000n$ iterations, while SA consistently performs slightly better than CA guided by coupling constants. It is worth noting, however, that the method of counting iterations (see Section~\ref{sec:Algorithmic_Design}) slightly disadvantages the CA in this comparison.

To explain these performance differences, we examine the role of frustration, a key factor in spin glasses where competing interactions prevent spins from aligning to minimize the energy of all interactions simultaneously. We quantify frustration using the misfit parameter from~\cite{Kobe1995}, defined for a state $\mathbf{x}$ with energy $E_\mathbf{x}$ as:
\begin{equation}
\mu_\mathbf{x} := \mu(E_\mathbf{x}) := \frac{E_\mathbf{x} - E_{\text{min}}^{\text{id}}}{E_{\text{max}}^{\text{id}} - E_{\text{min}}^{\text{id}}},
\end{equation}
where $E_{\text{min}}^{\text{id}}$ and $E_{\text{max}}^{\text{id}}$ are the ideal minimal and maximal energies, assuming all local contributions are minimized or maximized, respectively. Averaging over the $20$ graphs, we find $\Bar{\mu}_0^{3} \approx 0.29$ for the ground state of degree-3 graphs and $\Bar{\mu}_0^{20} \approx 0.42$ for degree-20 graphs, indicating higher frustration in the latter.

Frustration significantly impacts the effectiveness of the proposed CA since the coupling constants, which guide cluster formation, become less reliable as frustration increases. In graphs with high frustration, the CA’s decisions to add spins to clusters can become counterproductive: although adding a spin may appear beneficial from a local perspective, it may negatively affect global energy optimization. Specifically, such additions can reduce the number of cut edges that should ideally remain uncut -- or increase the number of uncut edges that should be cut -- in optimal solutions, although the decisions reduce the energy locally.

To further illustrate this behavior, one can consider the case of a tree graph, where the absence of loops eliminates frustration entirely. On a tree graph, coupling constants match optimal correlations, allowing the cluster building process to take only correct decisions, and thereby enabling rapid convergence to the global optimum. For 3-regular graphs, lower frustration allows coupling constants to provide effective guidance, leading to superior performance over SA. In contrast, the higher frustration in 20-regular graphs degrades coupling constants' accuracy, causing the CA guided by coupling constants to be no better than single-spin updates in SA. This interplay highlights the sensitivity of our approach to graph structure and frustration levels. In the following, we address these issues by enhancing the CA with more reliable information derived from the correlation sources described in Section~\ref{sec:Deriving_the_Correlations}.

\begin{figure}[htbp]
    \centering
    \input{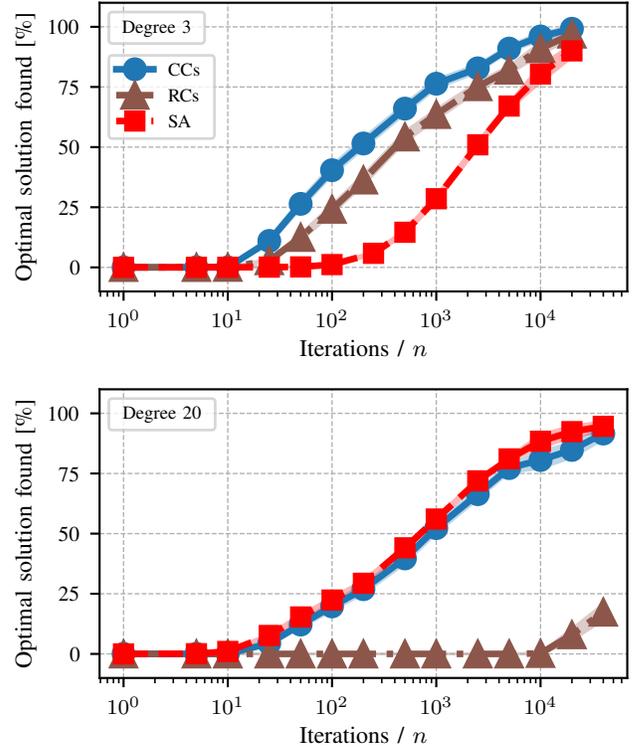}
    \caption{In the figure, the percentage of optimal solutions found is plotted against the number of iterations for twenty $3$-regular graphs (top) and twenty $20$-regular graphs (bottom) of size $n=100$. The standard deviation is shown as shaded area. While both coupling constants (CCs) and random clusters (RCs) outperform simulated annealing (SA) on 3-regular graphs (with a frustration witness of the ground state of $\bar{\mu}_0^3\approx0.29$), the higher frustration in $20$-regular graphs ($\bar{\mu}_0^{20}\approx0.42$) causes random clusters to fail in finding optimal solutions, even after $10{,}000n$ iterations, with SA slightly outperforming coupling constants.}
    \label{fig:CC_RC_SA}
\end{figure}

\subsection{Semidefinite Programming \& Monte Carlo Correlations}
\label{sec:semidefinite_programming_and_monte_carlo_correlations}

In this section, we analyze the CA guided by SDP and MC correlations for the same graphs as in Fig.~\ref{fig:CC_RC_SA}, in order to understand the effect of improving correlations and their interplay with increasing frustration on larger systems of size $n=100$, which are out of reach for QAOA simulations.

The left panels in Fig.~\ref{fig:MC_SDP_performance_and_corrs} present the performance of the CA, quantified by the number of optimal solutions found in dependency of the number of iterations, when guided by coupling constants, SDP correlations, and MC correlations sampled at different inverse temperatures $\beta_s$. The MC correlations are computed from $2{,}000$ bitstrings whose average solution quality is denoted as $\Bar{C}_{\text{app}}^{\text{MC}}$ and improves with increasing $\beta_s$. The quality of SDP correlations is evaluated by performing the GW hyperplane rounding procedure 1000 times and taking the average approximation ratio $\Bar{C}_{\text{app}}^{\text{GW}}$. For these plots, we set the scaling parameter to $\lambda_{\text{scale}}=1$ for simplicity. However, further hyperparameter tuning could enhance performance, particularly when employing correlations sampled at high inverse temperatures, where they become increasingly accurate. In such cases, higher $\lambda_{\text{scale}}$ leads to higher acceptance probabilities for adding spins to clusters, potentially improving the algorithm's effectiveness.

\begin{figure*}[htbp]
    \centering
    \subfloat[$3$-regular graphs]{%
        \input{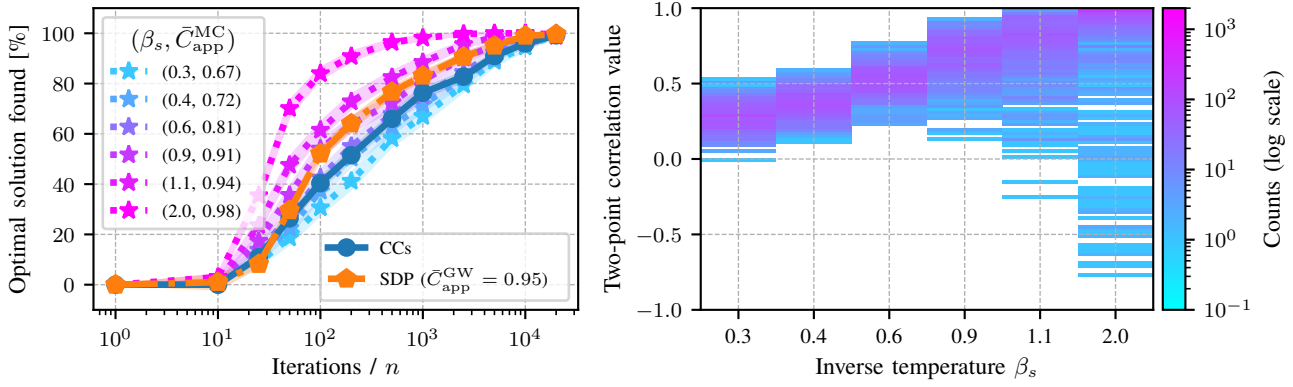}%
        \label{fig:MC_degree_3}%
    }\\[1ex]
    \subfloat[$20$-regular graphs]{%
        \input{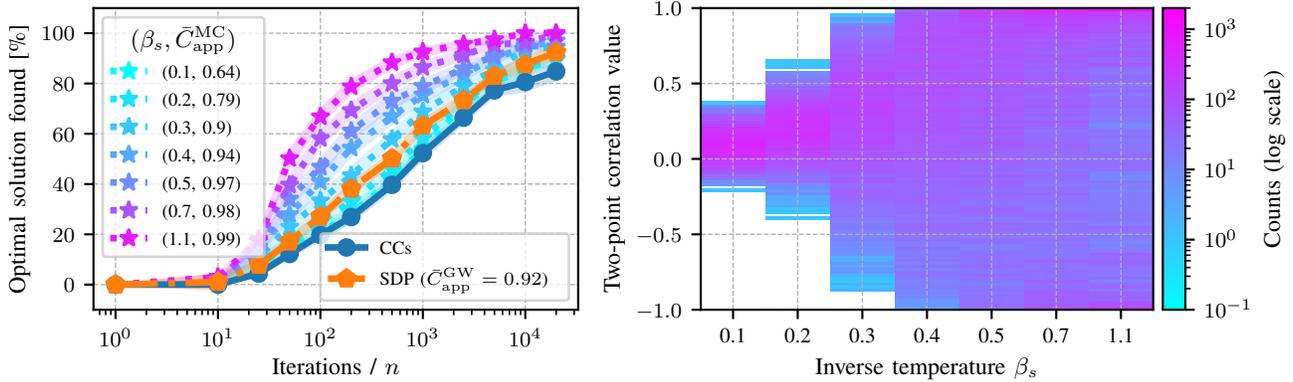}%
        \label{fig:MC_degree_20}%
    }
    \caption{The left plots display the number of optimal solutions found in percentage as a function of iterations for twenty $3$-regular graphs (a) and twenty $20$-regular graphs (b) of size $n=100$. The subfigures demonstrate the performance of the cluster algorithm (CA) guided by three different correlation types -- coupling constants (CCs), semidefinite programming (SDP) correlations, and thermal correlations sampled from Monte Carlo (MC) simulations at various inverse temperatures $\beta_s$. SDP correlations outperform the CA guided by coupling constants, yet they are surpassed by MC correlations of comparable solution quality, evaluated by the average approximation ratios $\Bar{C}_{\text{app}}^{\text{GW}}$ \& $\Bar{C}_{\text{app}}^{\text{MC}}$. On the right, the two-point correlation values $Z_{ij}^{\text{MC}}$ computed from MC sampling and sorted into bins are shown for the respective degrees for the same inverse temperatures as in the left plots. Note that only the correlations for edges with weight $J_{ij}=1$ are displayed, whose optimal correlations are $Z_{ij}^{\text{opt}}=1$ in the absence of frustration when minimizing the energy.}
    \label{fig:MC_SDP_performance_and_corrs}
\end{figure*}

We observe that the CA guided by SDP correlations consistently outperforms the CA guided by coupling constants (and consequently also SA) for both graph types. For 3-regular graphs, MC correlations sampled at $\beta_s=0.6$, corresponding to $\Bar{C}_{\text{app}}^{\text{MC}} \approx 0.81$, begin to slightly outperform coupling constants, and further improvements in correlation quality enhance the algorithm’s performance. In the case of 20-regular graphs, the positive effect of improved correlations is already visible at a sample temperature of $\beta_s=0.1$ (with $\Bar{C}_{\text{app}}^{\text{MC}} \approx 0.64$).

Furthermore, the CA guided by MC correlations appears to slightly outperform the CA guided by SDP correlations for comparable approximation ratios. Specifically, for 3-regular graphs, the performance at $(\beta_s, \Bar{C}_{\text{app}}^{\text{MC}}) \approx (1.1, 0.94)$ can be compared to the SDP correlation case where $\Bar{C}_{\text{app}}^{\text{GW}}\approx0.95$. Similarly, for 20-regular graphs, the MC correlation case at $(\beta_s, \Bar{C}_{\text{app}}^{\text{MC}}) \approx (0.3, 0.9)$ are comparable to the SDP correlation case where $\Bar{C}_{\text{app}}^{\text{GW}}\approx0.92$.

For future work, it would be interesting to investigate how this behavior scales with increasing system size and whether the performance gains achieved by using improved correlations outweigh the computational effort required to obtain them.

In the following, we explain the different improvements seen across the two graph types by looking at how much information the correlations carry and how this relates to the level of frustration in each graph. The right panels of Fig.~\ref{fig:MC_SDP_performance_and_corrs} display the correlation values $Z_{ij}^{\text{MC}}$ for all coupling constants with weights $J_{ij} = 1$, sorted into bins and plotted for various sample inverse temperatures for $3$- and $20$-regular graphs in (a) and (b), respectively. The data reveals that the correlation values for $3$-regular graphs increase in a linear fashion up to $\beta_s \approx 0.6$, after which they start to spread into the negative regime. This spreading behavior suggests that MC correlations sampled at higher inverse temperatures encapsulate more information about the frustration of the graph than coupling constants do. This is because, in low-energy states, frustration forces some edges to be cut, even though doing so locally increases the energy. By leveraging this increased information, the CA can make more informed decisions, effectively replacing the misleading guidance from frustrated coupling constants and thereby improving energy reduction in the cluster building process not only locally but also globally.

Therefore, around $\beta_s \approx 0.6$ we begin to observe performance improvements in the MC-enhanced CA compared to the coupling constants-guided version, as seen in the left panels of Fig.~\ref{fig:MC_SDP_performance_and_corrs}. Furthermore, for $20$-regular graphs, correlation values start spreading into the negative regime already at $\beta_s=0.1$, which coincides with the performance improvements of the MC-enhanced CA over the coupling constants-guided CA. As the inverse temperature increases further, the correlation values continue to spread across the full range of $[-1,1]$, indicating that they better resemble the structure of the low-energy space than coupling constants.

In summary, as the degree increases and frustration grows in low-energy states, coupling constants begin to provide misleading information to the cluster building process. This issue is mitigated by using low-energy correlations, which inherently encode a greater amount of information about frustration.

\subsection{Correlations Computed from QAOA}
\label{sec:Correlations_Computed_from_QAOA}

After thoroughly exploring the different effects that influence the algorithm's performance, we now turn to the main results. In this section, we examine the performance of the quantum-guided CA, where QAOA correlations -- computed as described in Section~\ref{sec:Deriving_the_Correlations} -- are used to guide the cluster building process.

To simulate QAOA up to graph sizes of $n=28$ and depths of $p=10$, we employ CUAOA, a GPU-accelerated QAOA framework~\cite{Stein2024}. This framework leverages GPU parallelization and the adjoint differentiation method to compute gradients efficiently.

Figure~\ref{fig:QGCA_depths} shows the performance of the CA guided by QAOA correlations obtained from circuits of varying depths, each yielding bitstrings with different average approximation ratios $\Bar{C}_{\text{app}}^{Q}$, as indicated in the plot, for twenty 10-regular graphs of size $n = 28$. Here, the scaling parameter $\lambda_{\text{scale}}$ was optimized in advance over integer values between $1$ and $10$, using $100$ repetitions for $50n$ iterations per value. The optimal scaling values were identified to be $\lambda_{\text{scale}} \in \{1, 6, 6, 6, 8, 8\}$ for the respective QAOA depths $p \in \{1, 2, 3, 5, 7, 10\}$. Notably, as the optimal performance shifts to values of $\lambda_{\text{scale}}$ greater than one, this indicates that the quality of the correlations improves strongly with increasing QAOA depth. In such cases, adding spins to clusters with higher probability proves to be highly beneficial for the algorithm’s effectiveness.
\begin{figure*}[htpb]
    \centering
    \subfloat[Quantum-guided CA for $10$-regular graphs for various iterations]{%
        \input{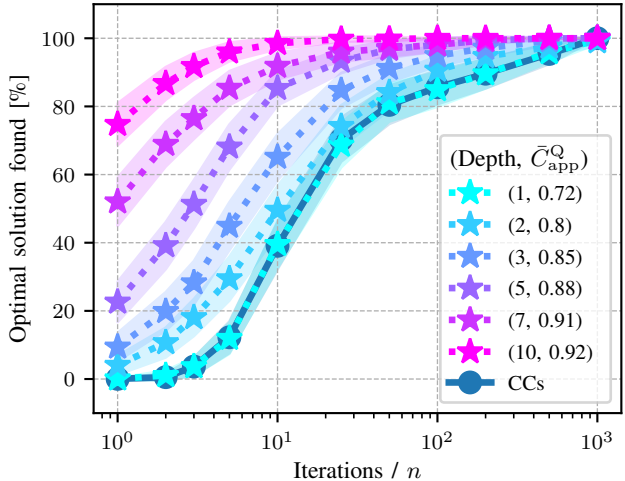}%
        \label{fig:QGCA_depths}%
    }
    \hfill
    \subfloat[Quantum-guided CA for graphs with various degrees for $500n$ iterations]{%
        \input{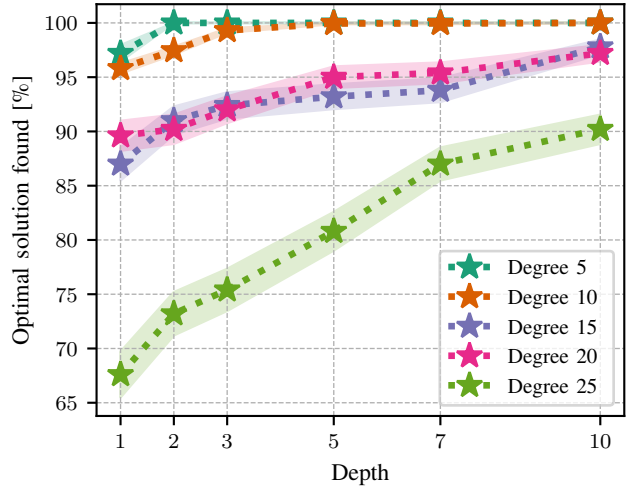}%
        \label{fig:qgca_degrees}%
    }
    \caption{In (a), the number of optimal solutions found in percentage as a function of iterations is shown for twenty $10$-regular graphs of size $n=28$. The cluster algorithm (CA) is guided by coupling constants (CCs) and correlations sampled from QAOA at different depths and therefore mean approximation ratios $\Bar{C}^{\text{Q}}_{\text{app}}$ of the sampled bitstrings. In (b), the CA is plotted for twenty regular graphs with various degrees, again of size $n=28$ and guided by correlations sampled from QAOA at the same depths as in (a).}
    \label{fig:QAOA}
\end{figure*}
Since SDP correlations are outperformed by coupling constants on these small graph instances, they are excluded from this analysis. SA exhibits performance comparable to the CA guided by coupling constants and is also not shown in the following plots. Lastly, it is important to mention that for certain values of $\lambda_{\text{scale}}$, we have observed a decrease in performance for increasing QAOA depths.

The performance curve for depth $p=1$ closely resembles that of the CA guided by coupling constants. In the following, we demonstrate why this is the case for $d$-regular graphs with edge weights drawn from $\{-1,1\}$ by utilizing the general analytical expression for the correlation values of QAOA at $p=1$, as derived in~\cite{Ozaeta2022}:
\begin{align}
    \begin{split}
        Z_{ij}^{p=1} &= \sin(2\beta_1)\cos(2\beta_1)\sin(2\gamma_1 J_{ij}) \\
        &\quad \times \Biggl[\prod_{\substack{k\neq i \\ k\neq j}}\cos(2\gamma_1 J_{ik}) 
        + \prod_{\substack{k\neq i \\ k\neq j}}\cos(2\gamma_1 J_{jk})\Biggr] \\
        &\quad - \frac{\sin(2\beta_1)}{2} \Biggl[\prod_{\substack{k\neq i \\ k\neq j}}\cos 2\gamma_1 (J_{ik}+J_{jk}) \Biggr. \\[1ex]
        &\qquad \Biggl.\qquad {}- \prod_{\substack{k\neq i \\ k\neq j}}\cos 2\gamma_1 (J_{jk}-J_{ik})\Biggr].
    \end{split}
\end{align}
Here, $(\beta_1, \gamma_1)$ denote the QAOA circuit parameters. Exploiting the symmetries of trigonometric functions, the first summand can be rewritten as $c_1 J_{ij}$, with $c_1 = 2\sin(2\beta_1)\cos(2\beta_1)\sin(2\gamma_1)\cos^{d-\delta_{1, \text{dist}(i,j)}}(2\gamma_1)$, since $J_{kl} \in \{-1, 0,1\}$. Here, $\delta_{1,\text{dist}(i,j)}$ denotes the Kronecker delta, which is one if $\text{dist}(i,j)=1$ and zero otherwise.

For the second summand, we distinguish between the two cases: $\text{dist}(i,j) > 2$ and $\text{dist}(i,j) \leq 2$. In the former case, the summand evaluates to zero, as both product terms contribute a factor of $\cos^{2d}(2\gamma_1)$. In the latter case, we must further differentiate between whether $i$ and $j$ are direct neighbors. If they are, the term introduces a factor of $\pm c_2\cos^{2d-1}(2\gamma_1)\left[\cos(4\gamma_1)-1\right]$ with $c_2 = -\frac{\sin^2(2\beta_1)}{2}$, where the sign depends on the coupling constants $J_{ik}$ \& $J_{jk}$. If $i$ and $j$ are not direct neighbors but still satisfy $\text{dist}(i,j) \leq 2$, the term evaluates to $\pm c_2 \cos^{2d-2}(2\gamma_1)\left[\cos(4\gamma_1)-1\right]$. Consequently, for $d$-regular graphs with edge weights drawn from $\{-1,1\}$, the correlation matrix for $i \neq j$ and $p=1$ is given by:
\begin{align}
    \begin{split}
        Z_{i\neq j}^{p=1} =\,& c_1 J_{ij} \\
        \pm \, & c_2 
        \begin{cases}
        \;\; 0, & \text{if } \mathrm{dist}(i,j)>2, \\[1ex]
        \begin{array}[b]{l}
        \cos^{2d-\mathrm{dist}(i,j)}(2\gamma_1)\\[1ex]
        \quad \times \Bigl[\cos(4\gamma_1)-1\Bigr],
        \end{array}
        & \text{otherwise.}
        \end{cases}
    \end{split}
    \label{eq:QAOA_p_1_equal_A}
\end{align}
For optimized parameters, the average ratio of the second to the first summand in~\eqref{eq:QAOA_p_1_equal_A} -- evaluated over all neighbors -- remains below $5.1\%$ across the twenty $10$-regular graphs analyzed in Fig.~\ref{fig:QGCA_depths}. Thus, the coupling constants are well-approximated by a scaled version of the correlations for neighboring nodes, which serve as the basis for cluster construction.

Furthermore, in~\cite{Dupont2024}, a statement made in the context of the Sherrington-Kirkpatrick model with random edge weights drawn from $\{-1,1\}$ suggests that, in the limit $n \rightarrow \infty$ and with optimal parameters, the two-point correlation matrix at $p=1$ is given by
\begin{equation}
    Z^{p=1}_{i\neq j} = \frac{1}{\sqrt{en}} \left(J_{ij} + K_{ij}\right),
\end{equation}
where $e$ is Euler's number and $K_{ij}$ follows a normal distribution with mean zero and variance $1/4e$. Consequently, in this case with optimal parameters, the correlation matrix $Z$ closely resembles the coupling constants, with a scaling factor and an additional summand of $\pm\frac{1}{2e}$.

Then, as expected, increasing QAOA depths -- leading to better approximation ratios $\Bar{C}_{\text{app}}^{Q}$ of the sampled bitstrings -- significantly enhance the performance of the quantum-guided CA. This result highlights the benefits of using quantum-derived correlations, as higher-depth QAOA circuits capture increasingly accurate problem structure information, which translates directly into better guidance for the CA.

In Fig.~\ref{fig:qgca_degrees}, we evaluate the performance of the quantum-guided CA across regular graphs of different degrees. The experiments are conducted on 20 graph instances of size $n=28$, using correlations computed from QAOA at various depths, for $500n$ iterations with $50$ repetitions per data point. The results indicate a clear trend: graphs with lower degrees are significantly easier to solve compared to those with higher degrees and thus higher frustration. Furthermore, increasing the QAOA depth leads to improved performance across all degrees. The hyperparameter $\lambda_{\text{scale}}$ has been optimized in the same manner as for the experiments shown in Fig.~\ref{fig:QGCA_depths}. Again, we have also observed certain values, with which performance decreases with higher depths.

Finally, in Fig.~\ref{fig:QAOA_acc_probs}, we plot the acceptance probabilities in SA and in the correlation-guided CA for flipping clusters built on coupling constants and QAOA correlations sampled from circuits of different depths. These results are obtained for running the CA over $500n$ iterations, with $50$ repetitions, using the same graph instances as those in Fig.~\ref{fig:QGCA_depths} and recording the acceptance probabilities when $\beta \in [1.0,8.0]$. We observe that the acceptance probabilities of SA are similar to those of the CA with coupling constants. This behavior is expected, as SA performs similar to this particular variant of the CA, as discussed earlier. Furthermore, the data shows that the quantum-guided CA using QAOA is highly efficient. At $p=1$, the acceptance probability remains similar to that of coupling constants, but already at $p=2$, the median increases to approximately $30\%$. Notably, for $p=10$, it rises to around $95\%$, indicating a strong improvement in cluster moves as QAOA depth increases.

\begin{figure*}[htbp]
    \centering
    \input{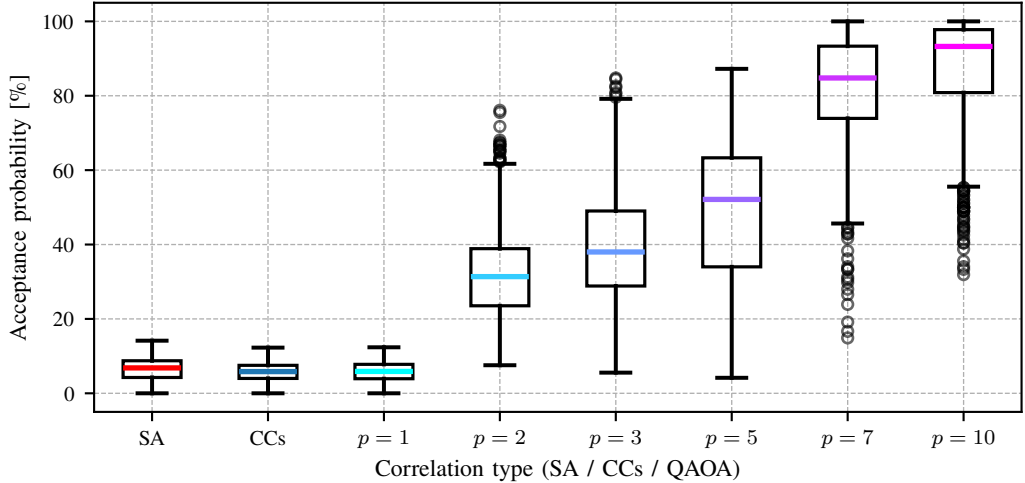}
    \caption{The acceptance probabilities of accepting cluster flips in the inverse temperature interval $\beta \in [1.0, 8.0]$ is shown for simulated annealing (SA), the cluster algorithm guided by coupling constants (CCs), and samples computed from QAOA with different depths $p$ for the same twenty $10$-regular graphs as in Fig.~\ref{fig:QGCA_depths} with $50$ repetitions with each $500n$ iterations.}
    \label{fig:QAOA_acc_probs}
\end{figure*}

\section{Conclusion}
\label{sec:Conclusion}

In this work, we propose a novel CA that leverages precomputed two-point correlations to enhance the optimization of combinatorial problems in the form of \textsc{Max-Cut}. It can be seen as a post-processing strategy, to refine solutions obtained by (quantum) approximate optimization algorithms, such as QAOA or SDP. By using these correlations to guide cluster formation, the algorithm enables large transitions in configuration space, helping to escape local minima where conventional single-spin update methods struggle. By employing low-energy correlations in the cluster building process, it overcomes these challenges of percolation, which hinders alternative algorithms from effectively exploring frustrated systems. We have analytically demonstrated the close relationship between QAOA correlations at $p=1$ and coupling constants, showing that they exhibit similar structural properties. As the QAOA circuit depth increases, the extracted correlations become increasingly refined, capturing more problem-specific information and potentially enhancing the algorithm’s performance. While the results for small systems are promising, further studies on larger graphs -- especially with high degrees -- are necessary to fully understand the scalability and practical advantages of QAOA correlations in the cluster building process.

By examining the CA guided by SDP and thermal correlations on larger systems beyond the reach of QAOA simulations, we highlight the crucial role of frustration in shaping the effectiveness of different correlation types. In systems with little frustration, where coupling constants already provide strong guidance, the improvements from higher-quality correlation matrices are less pronounced. However, in highly frustrated systems, where coupling constants alone provide misleading information, we observe significant benefits from more informative correlations.

There are multiple promising extensions for future work, such as modifying the cluster-building procedure to align more closely with Markov chain Monte Carlo methods, ensuring properties such as ergodicity and detailed balance, which would broaden its applicability. Furthermore, benchmarking the algorithm against correlations obtained from alternative quantum methods, such as Variational Quantum Eigensolvers or Quantum Annealing, could provide deeper insights into the comparative advantages of different approaches. Another important question is whether filtering correlations -- such as considering only the lowest-energy bitstrings -- could improve the quality of the guiding information. Moreover, the effects of noise in the Noisy Intermediate-Scale Quantum (NISQ) era should be analyzed.

Finally, a key question that remains is whether the computational effort required to compute high-quality correlations is justified by the speedup they provide in solving optimization problems. Understanding this trade-off, particularly as system size increases, will be crucial in determining the practical scalability of our approach. Future studies on larger graphs with higher degrees will be essential to evaluate whether the observed benefits persist in more challenging problem instances.

\section*{Acknowledgments}

P.J.E was partially funded by the German BMWK project QCHALLenge (Grant No. 01MQ22008B).

\bibliographystyle{IEEEtranDoi}  
\bibliography{bstcontrol,references}

\end{document}